# Two Lifshitz points in melt-crystallized polymers: nanostructure, necking, energy dissipation

A. N. Yakunin[*]

*Karpov Institute of Physical Chemistry, 103064 Moscow, Russia*

**Abstract:** In frameworks of scaling theory of phase transitions and critical phenomena the structure of melt-crystallized polymers is discussed. The model constructed follows drawing polymeric materials and dissipating the energy during the transition from isotropic to oriented state. It is possible to estimate temperatures of secondary relaxation transitions. The "entropic" parameters of the model are: the space dimension; the number of components of an ordering field; the polymerization degree. The Kuhn segment, the melting temperature and the difference of energies of two rotating isomers can be regarded as "energetic" parameters. The model enables to calculate a number of important structural and dynamic physical values: the density ratio of crystalline to amorphous phase; the thicknesses of crystalline and amorphous layers in an isotropic lamellar material; the crystallinity degree; the fluctuation spacing due to entanglements; the neck draw ratio; the draw ratio at break; the elastic deformation connected with *gosh-trans* transitions; etc. This model is based on the assumption that there are exist two Lifshitz points in such polymers. The neck draw ratio characterizes an irreversible process during which the energy dissipation is observed. It is the main difference from magnetic systems, for example. The results obtained are in a good agreement with experimental data including the results found for solution-crystallized polymers.

## INTRODUCTION

It has been revealed [1] that the neck draw ratio, $\lambda_n$, in linear melt-crystallized MC PE decreases with increasing $M_w$ as $\ln\lambda_n = A - \beta\ln M_w$ where A a positive constant and $\beta \approx 0.3$ is the critical exponent of the fluctuation theory of the second order phase transition. It has been also shown [1] that the transformation to the oriented state is the first order phase transition forming 3 stages: swelling the polymer under applied stress, dissolving crystallites and crystallizing extended macromolecules. Since upon acting an external field the second order phase transition has not a well-defined position [2] it is available to suppose that the melting temperature, $T_m$, is the Lifshitz point for MC polymers (let us recall that the lines of the 1-st order and the 2-nd order phase transitions intersect each other in this point [2]). However, where is the Lifshitz point in solution-crystallized (SC) polymers? We shall see below that the point can be attributed to the temperature of $\alpha$ – relaxation transition.

[*] E-mail: yakunin@cc.nifhi.ac.ru

The aim of the article is to find the minimum number of parameters which is necessary for the description of behaviour of concentrated polymer systems. We are going to show that at least there are the six parameters. The three "entropic" parameters are: the space dimension, d; the number of components of an ordering field, n; the polymerization degree, N. The Kuhn segment, $l_K$, the melting temperature, $T_m$, and the difference of energies of two rotating isomers, $\Delta E$, are the "energetic" parameters. We will see that a number of important parameters can not be calculated in frameworks of the mean-field approximation for the model of lattice gas since the latter theory does not take into account at all some physical phenomena such as breaking the symmetry and its recovery at large scales. Since an interest to problems mentioned above is great up to now [3-8] the present theory is suggested. The results obtained are compared with well-known experimental data, mainly, for linear high density PE (HDPE) due to a very wide available range of molecular masses (MMs) [9].

**RESULTS AND DISCUSSION**

**Two-phase model of polymer structure; an estimation of density ratio.** Let a polymer chain with the polymerization degree, N, walk over two lattices having a total boundary, one of them being the simple cubic. In that case z is the number of neighbour monomers surrounding a site of the lattice (z = 2d = 6 for three-dimensional space). If we will regard z* as the same parameter for the second lattice and introduce the probability p = z*/z < 1 then we can write the statistic sum, Z,

$$Z = \sum_{N=0}^{\infty} p^N = 1/(1-p)$$

where p can be defined as the probability discovering a monomer of the chain in the first phase while 1 - p is the same probability for the second phase. Assuming Z = z* we find

$$1 = Z(1-p) = pz(1-p) \qquad (1)$$

From the solution of the equation (1) we receive $p_{\pm} = 0.5(1 \pm (1 - 4/z)^{1/2})$ and $Z = 3 + 3^{1/2} \approx 4.732$:

$$Z/z \equiv p = p_+ = 1 - p_- = 1 - 1/Z.$$

The other solution $p_- = 1/Z = 1 - p$ has no interest for subsequent considerations.

Taking the logarithm of the expression (1) we obtain

$$\ln Z = \ln p + \ln z. \qquad (2)$$

Multiplying the equation (2) by $c/\ln z$ where $c$ is the mean concentration of chain monomers in the first phase we find

$$c \ln Z / \ln z = c \ln p / \ln z + c = C$$

where $C$ is a mean concentration. We suppose that $C$ is the concentration of monomers of the chain in the second phase. Thus, we can draw an important conclusion: for the two-phase model of polymer structure the density ratio of the densest phase to the second phase is the constant which is equal to

$$\ln z / \ln Z \approx 1.153$$

for three-dimensional space.

It should be underlined that for a number of semicrystalline polymers this relation is correct within the error which is less than 4%. In Table 1 the experimental data [10] are presented for 12 polymers. The list may be expanded to a great extent.

**Table 1.** The density ratio of the crystalline phase to the amorphous one for some polymers. The theoretical value is 1.153.

| polymer | the ratio densities, $\rho_c/\rho_a$ |
|---|---|
| polyethylene | 1.177 |
| polyester, $-(CH_2)_2-O-CO-(CH_2)_n-$, n = 8 | 1.167 |
| *trans*-1-4-polybutadiene | 1.163 |
| *trans*-1-4-poly-2-methylbutadiene, $\alpha$ - form | 1.161 |
| polyester, $-(CH_2)_2-O-CO-(CH_2)_n-$, n = 6 | 1.152 |
| polytetrafluoraethylene | 1.151 |
| polyoximethylene | 1.126 |
| polymeric selenium | 1.125 |
| *cis*-1-4-polybutadiene | 1.122 |
| polyvinyledenefluoride | 1.122 |
| polyester, $-(CH_2)_2-O-CO-(CH_2)_n-$, n = 4 | 1.111 |
| isotactic polypropylene | 1.110 |

The value of $Z \approx 4.732$ is very close to that of 4.68 which is a characteristic of the polymer statistics for the excluded volume problem [11]. This enables us to solve our problems in frameworks of the scaling theory of phase transitions and critical phenomena. This is the second important conclusion obtained by help of the present model.

**Folding polymer chains.** Let us direct the co-ordinate axis along the unit vector from the densest crystalline phase to amorphous phase normal to the surface separated one lattice from other. Let the origin of co-ordinate be on the surface. From the discussion above one can assume that an external force compresses the dense phase and expands the other. As a result the chemical potential of "crystalline" and "amorphous" monomer is $\mu_0 + fx$ where $\mu_0$ is the chemical potential of a monomer for one-phase system if

$$f = 0. \qquad (3)$$

The condition (3) defines a critical point.

We will assume below that the fluctuation attraction [12] of chain ends compresses the one-phase polymer coil in one direction [13]. Breaking the symmetry which is observed, for example, upon decreasing temperature results in appearing the lamellar supramolecular lattice, the latter is the most typical nanostructure for crystalline polymer materials. We are going to calculate the thicknesses of crystalline and amorphous phases and to find that they are much less than the macromolecule length. It means that the chain can just turn back to that phase from which it walked. The phenomenon for crystalline phase is called by folding.

**The statistic sum obtained by des Cloiseuaxs' method.** Let us define the scale length, $l_s$, which is proportional to the Kuhn segment $l_K \sim \exp(\Delta E/T)$

$$l_s = a \exp(T_\beta/T). \qquad (4)$$

Here, $a$ is the diameter of monomer (the lattice constant), $T_\beta$ is the temperature of $\beta$ - relaxation transition, $T$ is expressed in energy units.

Our qualitative speculations give an opportunity to estimate those parameters which we wrote about. The results are in a good accordance with experimental data [1, 14-16]. The rigorous theory will be suggested below using des Cloiseuaxs' method [17]. This one is described by P.-G. de Gennes [11] and we will follow him.

Let us determine the free energy of amorphous phase, $F_2$, as

$$-\ln Z(T) = F_2/T = -(\varepsilon_0 + T_\beta - \sigma a^3)/T. \qquad (5)$$

Here, the average stress, $\sigma$, is defined as

$$S \equiv \sigma a^3/T = <fx>/T = T_\beta (1 - T/T_m)/T \qquad (6)$$

and we sign $\sigma a^3/T$ by $S$. At the Lifshitz temperature $T = T_L = T_m$, consequently, $S = 0$ (6) and $f = 0$ (3). $\varepsilon_0$ is non essential parameter. The subscript 2 is referred to the second amorphous phase. It should be noted that in literature [2] $S < 0$ prior to the transition point (upon increasing temperature) and S > 0 after it. Our definition is more convenient for experimentation researchers as the yielding stress, for example, increases with decreasing temperature.

Analogously we can determine the free energy of crystalline phase, $F_1$, as

$$-\ln z = F_1/T = -(\varepsilon_0 + T_\beta)/T - S. \quad (7)$$

From (1), (5-7) we can obtain

$$-(F_2 - F_1)/T = -2S = \ln(Z(T)/z) = \ln p(T) \quad (8)$$

From (8) we can define the temperature, $T_\alpha < T_m$, at which the expression (8) is in accordance with (1)

$$T_\alpha = (1/T_m - \ln p/(2T_\beta))^{-1}. \quad (9)$$

Put another way, $p(T_\alpha) = p = (3 + 3^{1/2})/6 \approx 0.789$. $T_\alpha \sim \Delta E$ is the temperature of $\alpha$ – relaxation transition in linear PE and below we will call it by the second Lifshitz point for MC PE.

At $T = T_m$ the jump-like first order transition occurs upon decreasing temperature. The homogeneous polymer system forms the two phases. Also $S = 0$, $Z(T_m) = z$, and $p(T_m) = 1$. Thus, we go to notation of the second order phase transition if the parameter $S \to 0$.

Let us expand the statistic sum of our two-phase polymer system in the series by help of des Cloiseuaxs' method [11, 17]:

$$\exp(-\Delta\Omega/T) = Z(H)/Z(0) = 1 + (H/\tau)^2 \sum_{N=0}^{\infty} ((K/\tau)(Z(T)/z))^N N^{\gamma-1} \quad (10)$$

where $\Omega$ is the Landau free energy, $(H/\tau)^2$ is the factor corresponding to the interaction of two ends of the chain, $K/\tau$ is the factor describing the interaction of neighbour monomers surrounding a lattice site, $Z^N N^{\gamma-1}$ is the total number of self-avoiding walks, $\gamma$ is the critical exponent. Let

$$\tau = \tau_{cr} \exp(-S)/Z. \quad (11)$$

Here, $Z \approx 4.732$. Sometimes we use the same symbols for constants and functions. However, we will make necessary explanations if one can not understand the difference from the text.

Substituting (8) for $Z(T)$ ($Z(T) = z \exp(-2S)$) and (11) into (10) we may find the sum replacing it by the integral if $KZ = \tau_{cr}$ [11]:

$$Z(H)/Z(0) = 1 + (H/\tau)^2 \int_0^\infty \exp(-SN) N^{\gamma-1} dN = 1 + (H/\tau)^2 \Gamma(\gamma)/S^\gamma. \quad (12)$$

We define the variation of a magnetic free energy as [11]

$$\delta F_m/\tau = (\tau_0/\tau) S^{\nu d} (M/S^\beta)^2 - M H/\tau \quad (13)$$

and, by differentiating it with respect to M, find the minimum

$$\delta F_m/\tau|_{min} = -(H/\tau)^2/(4(\tau_0/\tau) S^\gamma) = -M^2 (\tau_0/\tau) S^\gamma. \quad (14)$$

Thus, $\tau$ is the temperature of a magnetic system connecting with polymer one, $\tau_0$ is an energetic parameter of the magnetic system, M is the magnetic moment, H/$\tau$ is the field, and we use Vidom - Kadanoff relation [11]

$$\gamma - 1 + (1 - \nu d) = -2\beta \quad (15)$$

upon obtaining (14). $\nu$ is the critical exponent of correlation radius. Defining the magnetic susceptibility from (13) and (14) as

$$\chi\tau = M/(H/\tau) = ((2\tau_0/\tau) S^\gamma)^{-1} \quad (16)$$

we see that it diverges at $S \to 0$ (at $T = T_L$). Below we are going to use the property of $\chi$ to calculate some polymer structure parameters.

We can define from (10) the volume fraction of the chain $\Phi_p = Q^{-1} (\partial \ln(Z(H)/Z(0)))/(\partial \ln H^2)$ and $2\Phi_p + \Phi = Q^{-1} (\partial \ln(Z(H)/Z(0)))/(\partial \ln(1/\tau))$ where

$$(2\Phi_p + \Phi)/\Phi_p \approx \Phi/\Phi_p = N = \gamma/S, \quad (17)$$
$$Q = N^{\nu d} = \Phi_p^{-1} \quad (18)$$

is the total number of cites of the lattices [11]. At last, in that case the osmotic pressure is determined as

$$Q\pi a^3/T = \ln(Z(H)/Z(0)). \quad (19)$$

**Definition of the crystallinity; nanostructures. Fracture of polymers.** Let us define the following structure parameters:

$$l_a = k\, l_K\, N^{\gamma-1}, \quad (20)$$
$$l_c = c(S)\, l_K, \quad (21)$$
$$y = l_a/l_c = (k/c(S))\, N^{\gamma-1} = v/(1-v), \quad (22)$$

where $l_a$ and $l_c$ are the average thicknesses of amorphous and crystalline layer, respectively, k is the constant and c(S) is the parameter we define below; v is the volume fraction of the amorphous phase which is equal to $l_a/(l_c + l_a)$ for the lamellar symmetry of our task.

We can find by differentiating with respect to S

$$-N^{-1}\partial v/\partial S = N^{-1}(1+y)^{-2}\, \partial y/\partial S = -v^2\, (\partial c(S)/\partial S) k^{-1} \gamma^{-\gamma}\, S^\gamma \quad (23)$$

where $v = y/(1+y)$ (see (22)) and we use (17). Supposing

$$k\, \gamma^\gamma = 1 \quad (24)$$

we can see that (23) is in accordance with (14) if

$$M^2 = v^2 \quad (25)$$

and

$$\partial c(S)/\partial S = \tau_0/\tau. \quad (26)$$

The subsequent study is possible using an analysis of pair correlation functions which has been obtained in [13].

Let us define the mean magnetic correlation [11]

$$\langle M(r_0)M(r)\rangle = (r_0/r)^{1+\eta} \qquad (27)$$

where η is the critical exponent of ordering field and the following relationship is true [11]:

$$2\beta = \nu(d - 2 + \eta).$$

Taking into account that at critical point the correlation radius [11]

$$\xi = aN^\nu \sim r \qquad (28)$$

we see that

$$\langle M(r_0)M(r)\rangle = (r_0/a)^{1+\eta} N^{-2\beta}. \qquad (29)$$

The definition (29) is in agreement with (19) since the integral of (29) over the total volume results in multiplying by Q from (18) and using (15) and (17) we obtain (16). Thus, the thermodynamic relationship (16) and the integral of the magnetic correlation (29) over the total volume confirm the validity of method.

It is reasonable to connect the mean magnetic correlation with $\lambda_n^2$. The neck draw ratio is a characteristic of ordering upon the orientation of semicrystalline polymers [1]. $\langle M(r_0)M(r)\rangle$ is a measure of magnetic ordering. Both $\lambda_n^2$ and $\langle M(r_0)M(r)\rangle$ have the same dependence on N.

It has been found [13] that the ordering parameter can be defined for the chain with excluded volume as

$$W^2 = (B/C) N^{-2\beta}$$

and $W_{cr}^2 = (B/C)^2$ at $N = N_{cr}$ where $B \approx 0.2068$ is the constant for Fisher's probability density, $C \approx 2842.45$, $C/B \approx 13744.9$, the number of components of ordering field n = 0 [18] for Wilson's ε – expansion [19]. If we define

$$\lambda_n^2 = W^2/W_{cr}^2 = (C/B) N^{-2\beta} \qquad (30)$$

and use the value of $\nu = 1-6^{-1/2}$, $\beta = 0.2998$ [1, 13] then $\ln N_{cr} \approx 15.89$. At $N = N_{cr}$ $\lambda_n = 1$ and it means that the oriented state cannot be observed. The conclusion agrees with the results of work [1]. See also Figure 1.

Taking the logarithm of (30) at $N = N_{cr}$ and dividing by ln(C/B) we obtain $0 = 1 - 2\beta \ln N_{cr}/\ln(C/B)$ and near $N_{cr}$ we can determine the value

$$K = 1 - \ln N/\ln N_{cr} \approx 1 - 0.0629\ln N. \qquad (31)$$

We call the K by crystallinity.

In general case both the crystallinity and some other parameters depend on the crystallization conditions. We assume here and below that the samples for investigations were prepared by help of standard techniques [1, 16].

Figure 2 shows that (31) is in accordance with experimental data [1, 16] up to high MMs (~ $10^6$ for HDPE) if to add some constant ~0.2 to the right hand of (31). The rest parameters can be calculated using (20) for the average thickness of amorphous layer, (21) and (22) for the average thickness of crystalline layer (K=1-v, Figure 3). If to interpolate $l_c$ by the equation $l_c = 32.79N^{-0.093}$ for K < 0.7 - 0.75 then at $N = N_{cr}$ $l_c^{cr} = 7.48$nm, $l_a^{cr} = 26.9$nm, naturally, K = 1- v ≈ 0.2. It is again the definition of the Lifhitz point since the first order phase transition occurs, the crystallinity changing from 0.2 to 0. At this point only an elastic deformation, $\lambda_{el}$, is possible. The elastic deformation is connected with *gosh-trans* transitions [1]. The results of [1] show that in oriented samples up to $\lambda_n$ at room temperature

$$\ln\lambda_{el} = 0.0428\ln(M_w/20000) \qquad (32)$$

where $N_{es}$ = 20000/14 ≈ 1430 is the number of monomers between entanglements of a secondary network (confirmed by SAXS data), 14kg/kmol is the MM of $CH_2$ group, see Fig. 1. This network has been revealed in [20]. The common determination differs from mentioned above. We can estimate the distance between ordinary entanglements from structural data as $l_c^{cr}$ = 7.48nm at $N = N_{cr}$. This is a minimum value of crystalline segment. The exponent 0.0428 is due to complex nature of viscous and elastic behaviour. It is connected with η [1, 13].

The fluctuation spacing due to the secondary entanglements [20] in high-oriented SC HDPE (draw ratio ~ 200) should be evaluated under additional assumptions: 1) the chain monomer has a double thickness owing to coupling, from this, 2) returning the self-avoiding walk to its beginning meets some difficulties due to the volume interactions [11]. We may write the following expression for the superlong space, L, using (20) and (24):

$$L = 2\gamma^{-\gamma}l_K(100/n_0)^{(\gamma-1)/\nu} N^{\gamma-1} \qquad (33)$$

where $n_0$ is the weight polymer concentration in solution which the sample was crystallized from. As shown in Fig. 4 this result is in a good agreement with experimental data [20] if $l_K$ = 2 nm and γ = 1.1757 from (15). If to assume that the relation (20) is connected with entanglements then at $n_0$ ≈ 18.5% $l_a = (l_K/2)(100/18.5)^{(\gamma-1)/\nu} N^{\gamma-1}$ where $l_K/2$ = 1 nm is the persistent length. Thus, MC HDPE seems to be prepared from 18.5%wt. solution hypothetically. It contains ~20% SC HDPE. In that case the polymer fracture upon drawing must correspond to spinodal mechanism.

Multiplying $\nu^{-1}\ln(r_0/r)$ by Vidom-Kadanoff relation (15) we can find the exact thermodynamic equation

$$<M(r_0)M(r)> = G_E(r) P(r)$$

where $G_E(r) = (r_0/r)^{d-1/\nu}$ is proportional to the Edwards correlation function, $P(r) = (r/r_0)^{(\gamma-1)/\nu}$ is the probability of collision of chain ends [11].

Multiplying $\ln(N/N_{cr})$ by Vidom-Kadanoff relation (15) we can see the thermodynamic identity

$$-F + G = -\Omega$$

where G is the Gibbs thermodynamic potential [2]. The singular part of free energy is due to the term $\ln N^{\gamma-1}$, the second term $\ln N^{1-\nu d}$ can be attributed to the chemical potential of dissolved substance (see (17) and (18)), at last the term from the right hand is proportional to the second derivative of $\Omega$ with respect to the ordering parameter [2].

Multiplying $\nu^{-1}\ln(n_0 r_0/100aN^\nu)$ by Vidom-Kadanoff relation (15) we can obtain the formulae

$$\lambda_n^2 = \lambda_{br}^2 l_a/l_a^{cr} \qquad (34)$$

where $n_0 r_0/100a = (C/B)^{1/(1+\eta)}$ (compare with (30)) and if $n_0 = 18.5\%$ then $r_0/a \approx 65631.24$ and

$$\lambda_{br}^2 = (n_0/100)^{d-1/\nu} N_{es}^2 N^{1-\nu d} \qquad (35)$$

$\lambda_{br}$ is the draw ratio at break, $(18.5/100)^{d/2-1/2\nu} \approx 0.33$, $N_{es} = (r_0/a)^{d/2-1/2\nu} \approx 1430.4$ is the number of monomers between entanglements of the secondary network, $l_a \approx 0.5\, l_K (100/n_0)^{(\gamma-1)/\nu} N^{\gamma-1}$. Let us recall that $l_a^{cr} \approx 0.5\, l_K (r_0/a)^{(\gamma-1)/\nu} \approx 26.9$nm is the value of $l_a$ at $N=N_{cr}$. These results are in accordance with experimental data [16] (see Fig. 5).

Only now we can begin to discuss the second Lifshitz point.

**The second Lifshitz point.** Let us define the temperature of $\beta$ – relaxation transition for SC polymers, $T_s$, from the proportion

$$T_m/T_\beta = T_\alpha/T_s. \qquad (36)$$

$T_s$ can be observed only by help of radiothermoluminescence (RTL) [21] (see Fig. 6). Insertion of a comonomer into the chain of low density polyethylene (LDPE) decreases the temperature of RTL maximum (about 180–185K in the case of LDPE). At $\beta$-maximum (240K for LDPE) unfreezing the mobility of molecules passing through amorphous layers of polymer occurs.

The results for LDPE [21], HDPE [22] and other polymers [9] in table 2 can be connected by means of proportion (36). $T_\alpha$ is close to the temperature of polymer solubility. About this temperature the relation (34) should be modified. The draw ratio at break shows a maximum [15, 23]. The lateral sizes of crystallites, $l_{110}$, in samples drawn up to $\lambda_n$ become more than the lateral sizes of crystallites in the same materials oriented up to break at $T > 373K$ [23]. It is a consequence that crystallites can not keep taut chains relaxing near $\alpha$ – transition. As a result, the draw ratio at break decreases upon subsequent increasing temperature. Sometimes, this process is called by melting. There is no reason for such terminology [1]. $T_\alpha$ is the second Lifshitz temperature for the part of SC HDPE in MC HDPE.

**Table 2.** The temperatures of phase and relaxation transitions [9] for some polymers. $T_s$ is calculated from (36). In contrast, for LDPE and HDPE $T_\alpha$ is estimated from (36).

| polymer | $T_m$, K | $T_\beta$, K | $T_\alpha$, K | T of solubility, K | $T_s$ |
|---|---|---|---|---|---|
| polyethylene (LDPE) | 388 [9] | 240 [21] | 299 (36) | > 341 [9] | 185 [21] |
| polyethylene (HDPE) | 407 [1] | 232 [22] | 335 (36) | > 353 [9] | 191 [22] |
| isotactic polypropylene | 459 | 275.5[i] | 385[ii] |  | 231 |
| polyvinyledenefluoride | 473 | 238[i] | 373[iii] | > 298 | 188 |
| Polyoximethylene | 448 | 198[i] |  | > 362 | 160[iv] |
| polybutene-1 | 415[v] | 252[i] | 335[vi] | > 373 | 203 |

[i] glass transition temperature

[ii] from $\Delta E$ [24] assuming $T_\alpha \approx \Delta E$ (see also (9))

[iii] $\alpha_1$ – transition

[iv] from T of solubility

[v] melting temperature of hexagonal phase

[vi] It is assumed that $T_\alpha$ has the same value as in HDPE

**The energy dissipation: macroscopic consideration.** At $N = N_{cr}$ the scale of correlations is $r_0 = a\ (C/B)^{1/(1+\eta)} > 10000a$ from (29) and (30), consequently, the macroscopic consideration is possible. Let us assume that the stress is the following function of time:

$$S(t) = S \exp(-t/t_M)$$

and

$$\partial S(t)/\partial t = - S/t_M \qquad (37)$$

where $S = \sigma a^3/T$ is the dimensionless (quasi)equilibrium value while $\sigma$ is the true stress applied to a polymeric sample, $t_M$ is the Maxwell relaxation time.

Let us determine the quadratic form characterized the energy dissipation [2] upon the plastic deformation of polymers

$$f = 0.5\ (S/t_M)\ (\lambda_n - 1)^2. \qquad (38)$$

The variation of the stress can be written from (37) and (38) as

$$\partial S(t)/\partial t = - (\lambda_n - 1)^{-1} \partial f/\partial (\lambda_n - 1) = - S/t_M.$$

The variation of entropy rate, $ds/dt$, is defined by the following equations [2]

$$ds/dt = (\partial s/\partial S(t))(\partial S(t)/\partial t) = - (\lambda_n - 1)^2\ (\partial S(t)/\partial t) = 2f. \qquad (39)$$

Thus, $\lambda_n$ is equal to 1 at the isotropic state as well as at the critical point [13]. Then from (39) $ds/dt = 0$. The entropy has a maximum at the equilibrium state. There is no drawing and, consequently, there is no dissipation of energy. The macroscopic consideration can serve as an model of the neck formation under irreversible deformations of polymers. At the critical point nothing change, but if $N \neq N_{cr}$ upon acting the stress the symmetry spontaneously breaks, and the polymer sample deforms through the neck formation.

The elastic modulus, E, can be expressed as the derivative of the thermodynamic potential with respect to the stress. Put another way, one can write using (39)

$$E = -(ds/dS) \approx -(ds/dt)/(\partial S(t)/\partial t) \sim (\lambda_n - 1)^2.$$

Upon the neck formation let us assume from (38)

$$S \rightarrow S(\lambda_n - 1), t_M \rightarrow t_M / (\lambda_n - 1).$$

Then

$$\eta \sim E\, t_M \sim (\lambda_n - 1),$$

$$v_{dr} \sim S/\eta = \text{const}$$

and one can see that the stationary state [25] with the draw velocity $v_{dr}$ = const is observed. Here, $\eta$ is the effective viscosity of system. The minimum of the energy dissipation takes place during drawing, the friction force depending on the draw velocity linearly [2].

**Conclusion remarks.** The relationships of parameters with MM are well-known [1, 14-16, 26].

WAXS and SAXS were used to study the changes in the density of amorphous regions an elastic stressing of oriented PE films of different MMs [26]. It was established that the value and sign of thermal effect of elastic deformation and the rule of variation of average density of amorphous regions are determined by the quantitative ratio of processes of condensation and thinning of amorphous ranges. It was shown that condensation of amorphous domains with elastic stressing takes place mainly in regions between fibrils, while thinning is in intrafibrillar regions. In proportion to the increase in MM of PE, the role of condensation of interfibrillar amorphous domains increases with elastic stressing of oriented samples.

The thermal effect of plastic deformation is defined by help of (39) [2]. Increasing the internal energy is due to the sum of work and released heat. The former depends linearly on $(\lambda_n - 1)$, the latter is quadratic in $(\lambda_n - 1)$. As a result the internal energy increases.

**Acknowledgements**. The author is thankful to Professor E.M. Antipov, to Professor S.N. Chvalun, to Professor I. Ya. Erukhimovich for helpful discussions, and to Dr. A. V. Mironov for technical help.

**REFERENCES**

1. A. N. Yakunin, *Intern. J. Polymeric Mater.,* **22**, 57(1993).

2. L.D. Landau and E.M. Lifshitz, *Statistical Physics,* 3rd ed., Pergamon, Oxford, (1980).

3. Y.-F. Yao, R. Graf, H.W. Spiess, D.R. Lippits, S. Rastogi, *Phys Rev E,* **76**(6), 060801(2007).

4. K. Shin, E. Woo, Y.G. Jeong, C. Kim, J. Huh, K.-W. Kim, *Macromolecules,* **40**, 6617(2007).

5. B.A.G. Schrauwen, L.C.A.v. Breemen, A.B. Spoelstra, L.E. Govaert, G.W.M. Peters, H.E.H. Meijer, *Macromolecules,* **37**, 8618(2004).

6. M.A. Monge, J.A. Diaz, R. Pareja, *Macromolecules,* **37**, 7223(2004).

7. Y. Song, K. Nitta, N. Nemoto, *Macromolecules,* **36**, 1955(2003).

8. P.G. Klein, M.A.N. Driver, *Macromolecules,* **35**, 6598(2002).

9. *Polymer Data Handbook,* Oxford University Press, (1999).

10. B. Wunderlich, *Macromolecular Physics, V.3,* Academic Press, New York and London, (1980).

11. P.-G. de Gennes, *Scaling concepts in polymer physics,* Cornell University Press, Ithaca and London, (1979).

12. A.Yu. Grosberg, A.R. Khokhlov, *Statistical Physics of Macromolecules,* American Institute of Physics Press, New York, (1994).

13. A.N. Yakunin, *Central Eur. J. Phys.,* **2**, 355(2003).

14. L. Mandelkern, *J. Phys. Chem.,* **75**, 3909(1971).

15. I.M. Ward in *Ultra-High Modules Polymers,* A. Ciferri and I.M. Ward, Eds., Applied Science, London, (1977).

16. R. Popli, L. Mandelkern, *J. Polym. Sci.: Part B: Polym. Phys.,* **25**, 441(1987).

17. J. des Cloiseuaxs, *J. Phys. (Paris),* **36**, 281(1975).

18. P.-G. de Gennes, *Phys. Lett.,* **38A**, 339(1972).

19. K.G. Wilson, J. Kogut, *Phys. Rep.,* **12C**, 75(1974).

20. S.N. Chvalun, A.B. Poshastenkova, N.F. Bakeev, *Polymer Science U.S.S.R.,* **34**, 161(1992).

21. S.M. Samoilov, V.A. Aulov, *Polymer Science U.S.S.R.,* **18**, 1124(1976).

22. A.N. Yakunin, A.N. Ozerin, S.G. Prutchenko, V.A. Aulov, N.I. Ivantscheva, O.V. Smolyanova, L.L. Spevak, N.F. Bakeyev, *Vysokomolek. soed.,* **B32**, 533(1990).

23. A.N. Yakunin, A.N. Ozerin, N.I. Ivantscheva, O.V. Smolyanova, A.V. Rebrov, L.L. Spevak, S.S. Ivantschev, V.S. Schirets, N.F. Bakeyev, *Polymer Science U.S.S.R.,* **30**, 854(1988).

24. J. O'Reilly, *J. Appl. Phys.,* **48**, 4043(1977).

25. P. Glansdorff, I. Prigogine, *Thermodynamic Theory of Structure, Stability and Fluctuations,* John Wiley & Sons, Inc., New York, (1971).

26. S.N. Chvalun, A.N. Ozerin, Yu.A. Zubov, Yu.K. Godovskii, N.F. Bakeyev, A.A. Baulin, *Polymer Science U.S.S.R.*, **23**, 1529(1981).

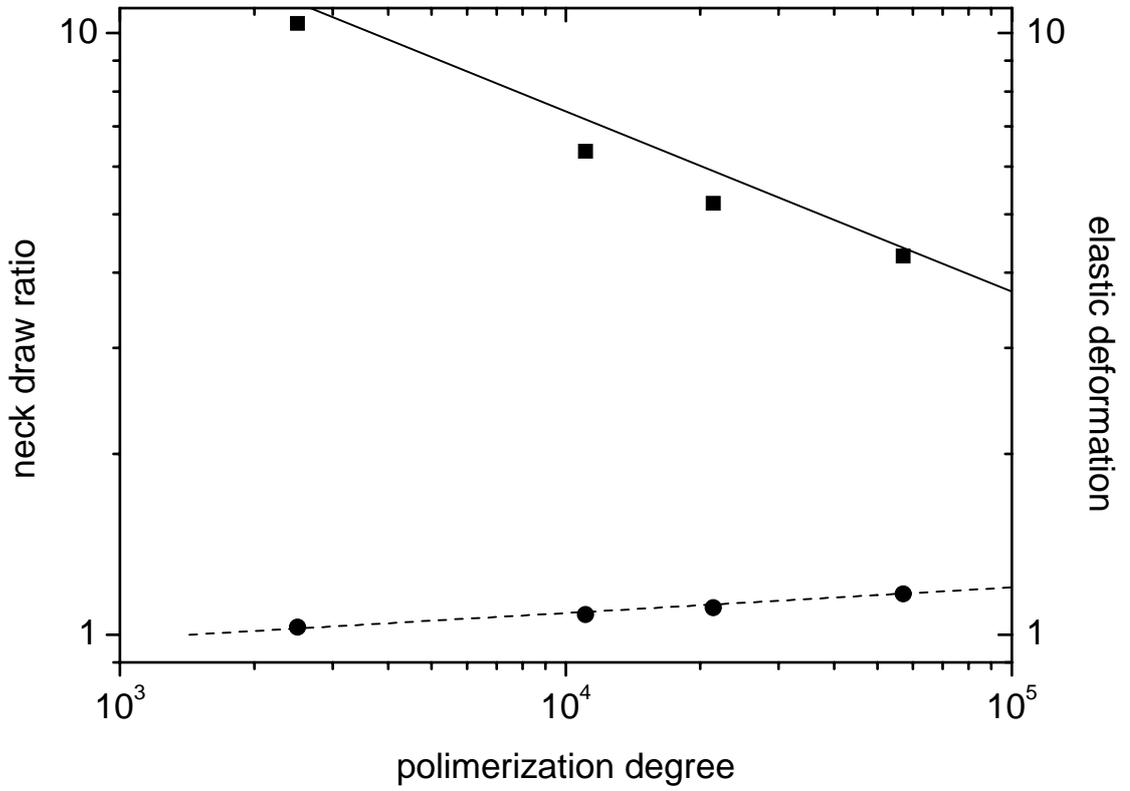

Fig. 1. The neck draw ratio, $\lambda_n$, *vs* the polymerization degree, N. Here and below $N = M_w/14$ where 14 kg/kmol is the MM of $CH_2$ group. Experimental data [1] are marked by squares, the theoretical solid curve $\lambda_n = 117.24\exp(-0.2998\ln N)$ corresponds to (30). The elastic deformation, $\lambda_{el}$, *vs* the polymerization degree, N. Experimental data [1] are marked by circles, the theoretical dash curve $\lambda_{el} = \exp(0.0428\ln(N/1430.4))$ corresponds to (32) where $N_{es} \approx 1430$ is the number of monomers between entanglements of secondary network.

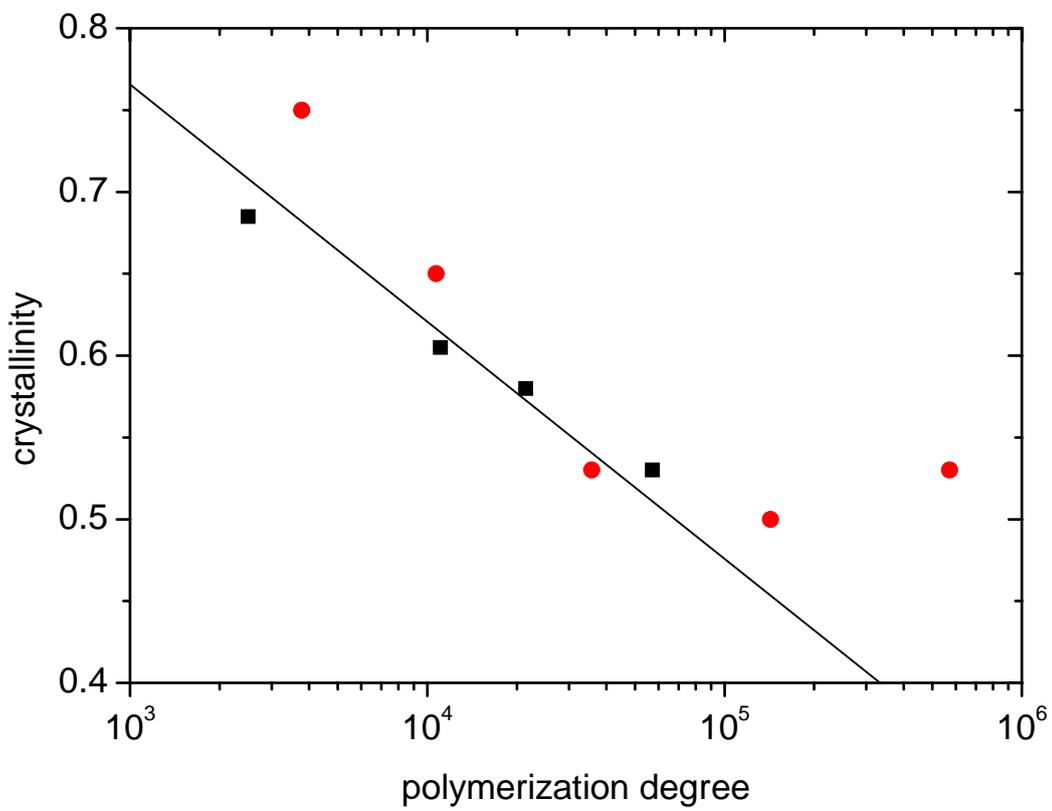

Fig. 2. The crystallinity degree, K, *vs* the polymerization degree, N. Experimental data [1] and [16] are marked by squares and circles, respectively, the theoretical curve K - 0.2 = 1 - 0.0629lnN corresponds to (31) where 0.2 is the volume part of SC HDPE in MC HDPE.

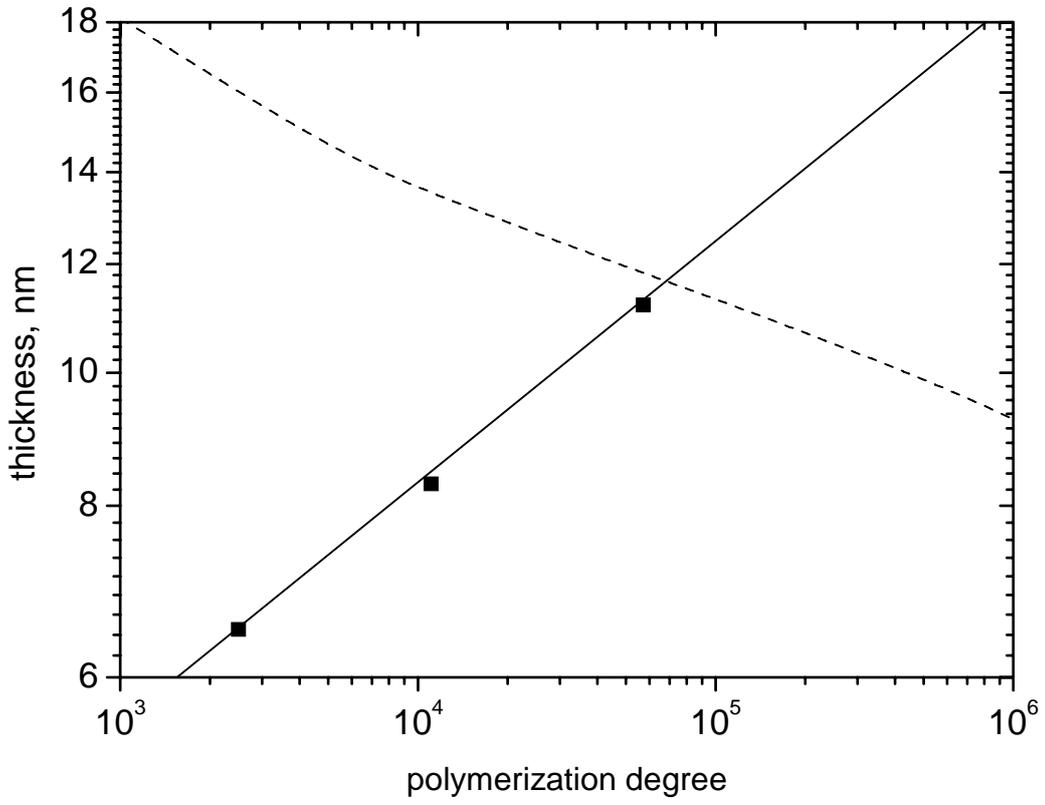

Fig. 3. The thickness of amorphous layer, $l_a$, *vs* the polymerization degree, N (solid line). Experimental data [1] are marked by symbols, the theoretical curve $l_a = 1.65\exp(0.1757\ln N)$ corresponds to (20) if $l_K = 2$ nm and $\gamma = 1.1757$ from (15). The theoretical thickness of crystalline layer, $l_c$, *vs* the polymerization degree, N (dash line).

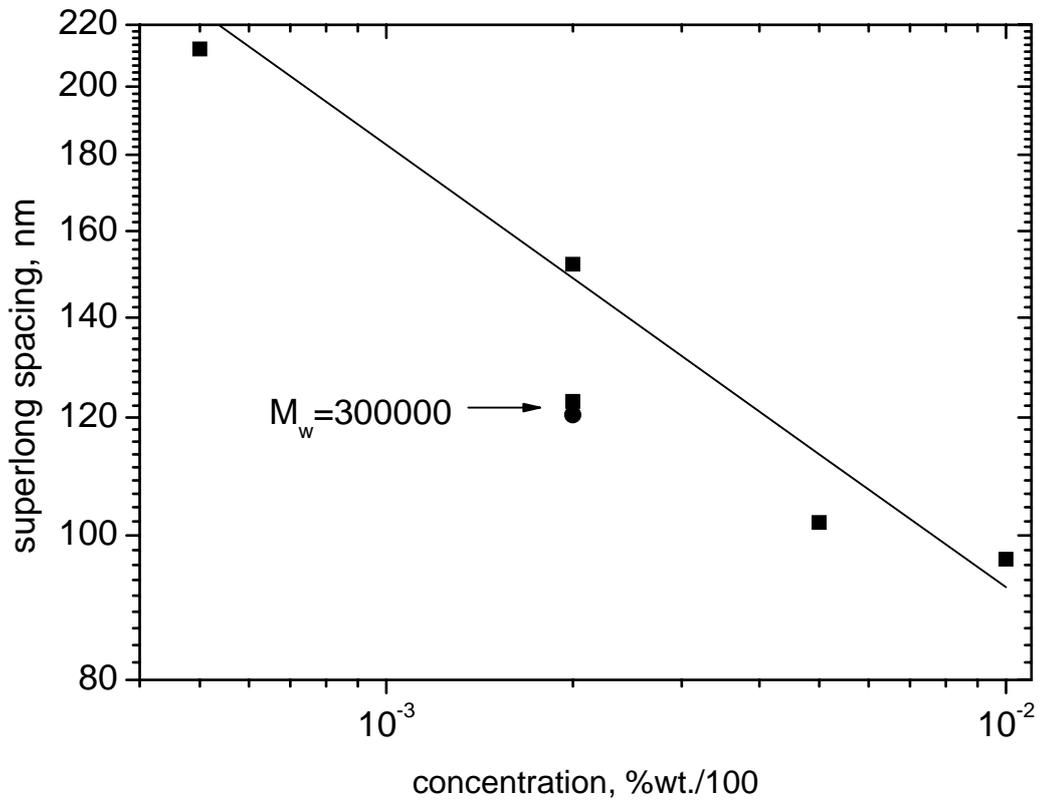

Fig. 4. The superlong spacing, L, *vs* the weight concentration, $n_0$, expressed in %. Experimental data [20] are marked by squares, the theoretical curve $L = 3.3\exp(0.2969\ln(100/n_0))\exp(0.1757\ln N)$ for $M_w = 1000000$ kg/kmol and the circle for $M_w = 300000$ kg/kmol correspond to (33) if $l_K = 2$ nm and $\gamma = 1.1757$ from (15).

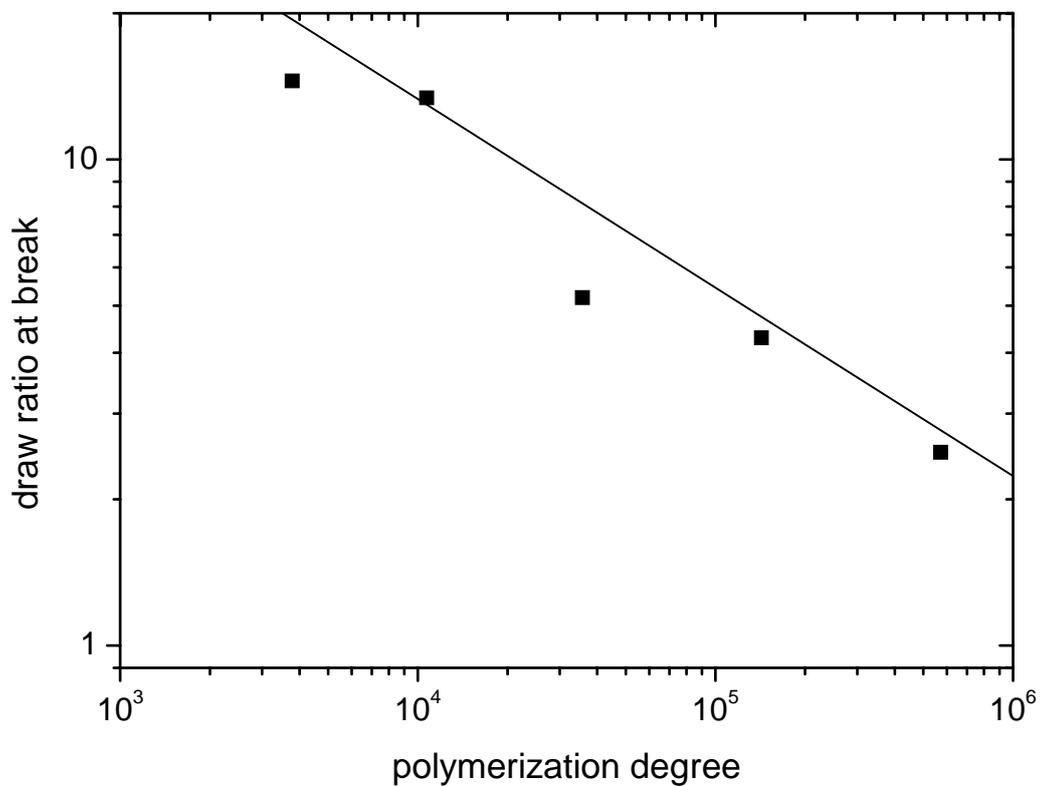

Fig. 5. The draw ratio at break, $\lambda_{br}$, *vs* the polymerization degree, N. Experimental data [16] are marked by squares, the theoretical curve $\lambda_{br} = 473.6\exp(-0.3876\ln N)$ corresponds to (35).

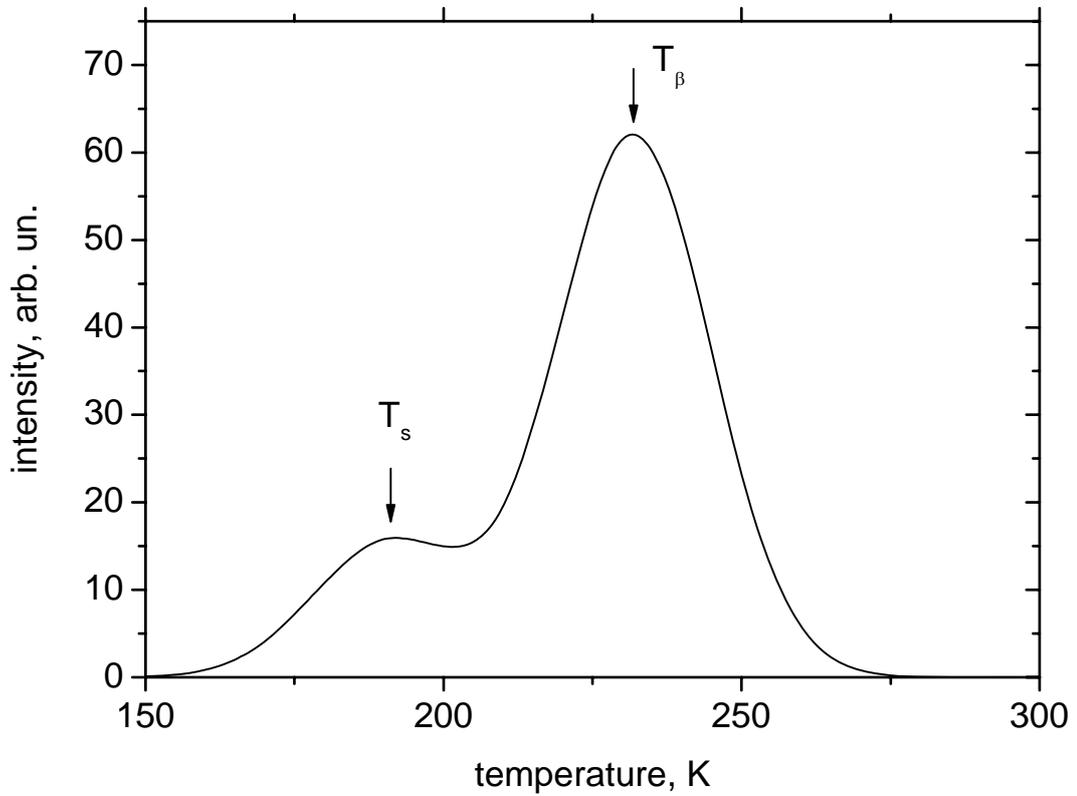

Fig. 6. The theoretical RTL curve for MC HDPE about its β – maximum shows $T_β$ and $T_s$ and agrees with experimental data [22]. Shifting to appropriate temperature an analogous curve of tangent of mechanical loss-angle for MC HDPE about its α – maximum will show $T_m$ and $T_α$ substituted for $T_β$ and $T_s$, respectively.